\documentclass[11pt,preprint]{aastex}

\newcommand{\kps}{\,\textstyle\rm{km~s}^{-1}}

\newcommand{\vlsr}{v_{\rm LSR}}

\slugcomment{Accepted AJ, 2015.02.02}
\shorttitle{GBT 4mm Survey of Orion-KL}
\shortauthors{Frayer et al.}

\begin{document}

\title{The GBT 67 -- 93.6 GHz Spectral Line Survey of Orion-KL}

\author{D. T. Frayer\altaffilmark{1}, 
Ronald J. Maddalena\altaffilmark{1},
M. Meijer\altaffilmark{1,2},
L. Hough\altaffilmark{1,3},
S. White\altaffilmark{1},
R. Norrod\altaffilmark{1},
G. Watts\altaffilmark{1},
M. Stennes\altaffilmark{1},
R. Simon\altaffilmark{1},
D. Woody\altaffilmark{1},
S. Srikanth\altaffilmark{4},
M. Pospieszalski\altaffilmark{4},
E. Bryerton\altaffilmark{4},
M. Whitehead\altaffilmark{1},
P. Ford\altaffilmark{1},
M. Mello\altaffilmark{1},
M. Bloss\altaffilmark{1}
}

\altaffiltext{1}{National Radio Astronomy Observatory, PO Box 2, Green
  Bank, WV 24944, USA}

\altaffiltext{2}{Eleanor Roosevelt HS, 7447 Cleveland Ave, Coronoa, CA, 92880, USA}

\altaffiltext{3}{Department of Physics, West Virginia University, P.O. Box 6315, Morgantown, WV 26506, USA}

\altaffiltext{4}{National Radio Astronomy Observatory Central 
  Development Lab, 1180 Boxwood Estate Road, Charlottesville, VA 22903, USA}

\begin{abstract}

  We present a 67--93.6 GHz spectral line survey of Orion-KL with the
  new 4\,mm Receiver on the Green Bank Telescope (GBT).  The survey
  reaches unprecedented depths and covers the low-frequency end of the
  3\,mm atmospheric window which has been relatively unexplored
  previously.  The entire spectral-line survey is published
  electronically for general use by the astronomical community.  The
  calibration and performance of 4\,mm Receiver on the GBT is also
  summarized.

\end{abstract}


\section{Introduction}

The Orion Nebula, the nearest region containing high-mass star
formation (at a distance of only 414\,pc, Menten et al. 2007), is one
of the most studied objects in the entire sky.  Within the Orion
Molecular Cloud 1 (OMC1) complex, the Kleinmann-Low (KL) nebular
harbors one of the most luminous embedded infrared sources
(Wynn-Williams et al. 1984).  The hot core and the intense shocks
associated with the ongoing star formation within the massive
molecular cloud give rise to very rich and complex chemistry (e.g.,
Esplugues et al. 2014).

Orion has been an important target of several spectral-line surveys
(e.g., Sutton et al. 1985; Turner 1989; Lee et al. 2001; Schilke et
al. 1997, 2001; Lee \& Cho 2002; Lerate et al. 2006; Olofsson, A. et
al. 2007; Carter et al. 2012), as spectral-line observations are key
to disentangling the chemistry and physical processes in star-forming
regions.  Spectral surveys at 1\,mm and shorter wavelengths are
complicated by line confusion.  Within the 3\,mm atmospheric window,
astronomical line confusion is less of an issue, and the window
contains the fundamental ground-state transitions of several key
molecular species.

The 4\,mm Receiver was commissioned on the Robert C. Byrd Green Bank
Telescope (GBT) in 2012.  In this paper, we present the 67--93.6 GHz
spectral line survey of Orion-KL taken with the new instrument.

\section{The 4\,mm Receiver}

The 4\,mm Receiver is a dual-beam, corrugated feed horn receiver that
covers the frequency range of approximately 67--93 GHz.  The system is
a single-side band receiver, and both feeds are within the cold
cryostat (15\,K).  The first version of the 4\,mm Receiver (White et
al. 2012) used a modified version of the W-band low-noise amplifiers
(LNAs) originally built for the Wilkinson Microwave Anisotropy Probe
(Pospieszalski et al.  2000).  This receiver exhibited noise
temperatures better than 80\,K over much of the frequency band and was
used during the initial observing season (through the summer of 2012).
In the fall of 2012, the receiver was upgraded with LNAs developed for
this band with more modern InP high-electron-mobility transistor
(HEMT) devices (Pospieszalski 2012, Bryerton at al. 2013).  The
upgraded amplifiers reduced the receiver noise to 40\,K over much of
the band.

Unlike other receivers on the GBT, there are no noise diodes for
calibration with the 4\,mm Receiver.  The receiver has an optical
wheel external to the cryostat that is used for calibration.  The
wheel also contains a quarter-wave plate for the conversion of linear
polarization into circular polarization for very long baseline
interferometry (VLBI) observations.  For calibration, the wheel has an
ambient temperature load ($T_{\rm amb}$ that is monitored by a sensor)
and an offset mirror used to view an internal cold load that has a
stable effective temperature ($T_{\rm cold}$), which is derived from
laboratory measurements.  For calibration, the table is rotated to
place the cold and ambient loads into each of the beams with
alternating observations.  Using the known temperatures of the loads,
the effective gains ($g$ [Kelvin/Volt]) of the system are measured.

\section{Observations}

Orion-KL was observed using the 4\,mm Receiver on the GBT (project
GBT12A\_364) during four observational sessions between 2012 May 30
and 2012 June 15.  The observed position of Orion-KL was $\alpha {\rm
  (J2000.0)} = 05^{\rm h}35^{\rm m}14\fs28$, $\delta{\rm (J2000.0)} =
-05\arcdeg22\arcmin27\farcs5$, corresponding to the bright continuum
peak (Fig. 1).  We adopt a source LSR velocity of $V_{LSR} = 8.8\kps$
(Turner 1989).  The observations were taken in position-switched mode
with alternating 1 minute ON and 1 minute OFF scans, using beam-1 of
the receiver with both linear polarization channels.  The OFF position
was offset by one degree in right ascension and at the same
declination as Orion-KL.

Twelve individual setups using the wide-bandwidth mode of the GBT
Spectrometer (see GBT proposer's guide) were used to carry out the
survey with a channel resolution of 0.39 MHz ($\sim 1.5 \kps$).  In
total, 38 overlapping 800 MHz spectrometer windows separated by 700
MHz were used to cover the full frequency range of 67.0 GHz to 93.6
GHz.  For each frequency set up, a sequence of 10 ON-OFF pairs of
Orion-KL observations were carried out along with short calibration
scans of the cold and ambient loads before and after the target
observations.  The observations for each setup were repeated twice,
resulting in a nominal total integration time of 20 minutes of
on-source data per frequency (20 ON-OFF pairs).

Observations of Jupiter and the nearby quasar 0423$-$0120 were used to
derive pointing and focus corrections every hour.  Representative
daytime pointing offsets during the observing sessions were
$2\farcs7\pm2\farcs1$ which is a smaller than the 10-13$\arcsec$ beam
and much smaller than the Orion complex which shows strong emission
over 1-3 arc-minutes.

Given the extended source size, most of the observations were done
without the ``AutoOOF'' measurements that are used to provide the
thermal corrections to the active surface (see GBT documentation for a
description of AutoOOF).  In contrast, optimal point-source
observations should be carried out with regular AutoOOF measurements
during the nighttime when the thermal stability of the dish is best.
Based on commissioning tests with the 4\,mm Receiver, the AutoOOF
corrections improve the point-source aperture efficiency by 30\% or
more.  Application of these corrections during the day are not
typically practical for the 4\,mm Receiver given that the thermal
environment of the dish is generally not sufficiently stable.  During
the day, the measured beam size was found to vary significantly (e.g.,
10--$13\arcsec$), but the beam shape remained well-behaved (fairly
symmetric and Gaussian).  Although the variation of beam size has a
direct impact on the point-source aperture efficiency ($\eta_{a}$), it
has little impact on the effective main-beam efficiency ($\eta_{mb}$)
used for the calibration of extended sources.  For example during the
commissioning of the instrument, we measured a beam size of
$10\farcs8$ at 77 GHz and derived $\eta_{a} = 31$\% and
$\eta_{mb}=50$\% in good nighttime conditions with the application of
the thermal AutoOOF surface corrections.  Without the AutoOOF
corrections, the beam-size increased to 12$\farcs5$ and the aperture
efficiency decreased to 23\%, but the main-beam efficiency remained
nearly constant at about 50\%.  Given the insensitivity of $\eta_{mb}$
for extended source emission, we adopt the main-beam temperature scale
$T_{mb}$ for the calibration of the data.

The main-beam temperature $T_{mb}$ is computed from the observed antenna
temperature $T_{A}$ using the relationship
\begin{equation}
T_{mb} = T_{A} \exp(\tau_{o}A)/\eta_{mb},
\end{equation}
where $\tau_{o}$ is the zenith opacity, $A$ is the airmass, and
$\eta_{mb}$ is the main-beam efficiency.  The opacity is derived from
the weather database for Green Bank which successfully forecasts the
opacity as a function of frequency and time with an accuracy of
$\Delta(\tau)= 0.006$ (Maddalena 2010a).  For the four observational
sessions, the zenith opacity varied only slightly ($\Delta \tau \simeq
0.02$--0.03).  The opacity variations contribute less than 3\% to the
total error budget of $T_{mb}$.

The antenna temperature for each ON-OFF scan pair is given by
\begin{equation}
T_{A} = T_{\rm sys} (V_{ON} - V_{OFF})/V_{OFF},
\end{equation}
where $T_{\rm sys}$ is the single-side-band system temperature and
$V_{ON}$ and $V_{OFF}$ are the measured voltages of the ON and OFF
scans.  The calibration scans are used to measure the gain ($g$) and
compute the system temperature for the OFF scan ($T_{\rm sys} =
g\,V_{OFF}$).  Based on the known temperatures ($T$) and measured
voltages ($V$) of the ambient ($amb$) and cold loads, the gain is
given by
\begin{equation}
g = [(T_{amb}-T_{cold})/(V_{amb} - V_{cold})].
\end{equation}

We adopt the median value of $T_{\rm sys}$ across each spectral window
in calibrating the spectra using equation (2).  This is referred to
``scalar'' calibration within GBT documentation and contrasts to
``vector'' calibration which uses $T_{\rm sys}$ computed as a function
frequency.  Currently for the 800\,MHz spectral windows, scalar
calibration gives significantly better baseline performance than
vector calibration for the 4\,mm Receiver.

Assuming a Gaussian beam, the main-beam efficiency is given by
\begin{equation}
\eta_{mb} = 0.8899\eta_{a}(\theta_{FWHM} D/\lambda)^2,
\end{equation}
where $\eta_{a}$ is the aperture efficiency, $\theta_{FWHM}$ is the
full-width half-maximum beam size in radians, D is the diameter of the
GBT (100m), and $\lambda$ is the observed wavelength (Maddalena
2010b).  The applicability of this equation was confirmed directly by
integrating the intensity profiles taken during the pointing scans.
Based on measurements as a function of frequency, $\theta_{FWHM}
\simeq 10\farcs8 ({\rm 77\,GHz}/\nu)$, where $\nu$ is the frequency in
GHz.

Observations of 3C279 in 2012 March were used to measure the aperture
efficiency.  For the GBT, $\eta_{a} = 0.352 T_{A}
\exp(\tau_{o}A)/S_{\nu}$.   Based on available measurements from the
Atacama Large Millimeter Array (ALMA) and the Combined Array for
Research Millimeter-wave Astronomy (CARMA), the derived flux density
for 3C279 at the time of the GBT observations was $S_{\nu}{(\rm
  77\,GHz)} = 23 \pm 2$\,Jy.  The corresponding derived aperture
efficiency at 77 GHz is $31\pm4$\%.  This efficiency is consistent
with net surface errors of $\epsilon = 280 \mu\rm{m}$ using the Ruze
equation:
\begin{equation}
\eta_{a} =0.71 \exp[-(4\pi\epsilon/\lambda)^2)],
\end{equation}
where the coefficient 0.71 is the aperture efficiency at long
wavelengths for the GBT.  Adopting $\epsilon = 280 \mu\rm{m}$,
equation (5) was used to compute $\eta_{a}$ as a function of
wavelength.

The calibration results based on 3C279 at 77 GHz were confirmed within
measurement errors with additional observations of Mars at 77 GHz
(2012 March), 3C84 at 89 GHz (2013 December), and 0927+392 at 89 GHz
(2013 March).  In terms of absolute flux calibration, there is some
degeneracy between the uncertainty of the temperature of the cold load
and the derived aperature efficiency, and hence $T_{mb}$.  The
effective temperature of the the cold load derived in the lab is known
to within 10\%, which corresponds to a 2.5\% uncertainty in $T_{mb}$.
The uncertainty in the cold load contributes negligibly to the total
error budget in comparison to the measurement errors of the
astronomical observations.  Factoring all known sources of errors, the
estimated uncertainty on the calibrated $T_{mb}$ temperature scale is
17\%.

\section{Results}

The observational parameters as a function of frequency are given in
Table~1.  The measured system temperature, aperture efficiency, and
main-beam efficiency are shown in Figure~2.  After data editing,
removing a low-order polynomial baseline, and combining the data for
both polarizations and observing sessions, the resulting empirical
noise level per 0.39\,MHz channel is 20--40\,mK over most of the
frequency range.  The 38 calibrated overlapping spectrometer windows
were spliced together to form a data set comprising of 68,096
frequencies and $T_{mb}$ values (Table~2).  The complete tabulated
data set is available as supplemental online material from the
journal.  Figure~3 displays the entire data set ploted on a logrithmic
scale.  To show the typical noise properties and baseline quality of
the data, an example portion of the survey is plotted as Figure~4.
Overlapping spectra separated by 0.5 GHz for the entire survey are
provided in the supplemental online material (Fig. 5).

The data are fairly clean, but a few artifacts remain in the data from
atmospheric O$_2$ and radio-frequency interference (RFI).  Table~3
lists the frequency of data features that are not associated with
Orion-KL.  The ``P-Cygni'' profiles from O$_2$ result from
position-switching.  Separate frequency-switched observations were
carried out to verify the association of these features with broad
atmospheric O$_2$ lines.

The detected spectral lines were fitted with a Gaussian using the
GBTIDL {\em fitgauss} routine.  Below 1\,K the data are very rich, and
the identifications of weak features become challenging in many cases
due to blended transitions and the associated with multiple possible
species.  In Table~4, we provide the results of fitting the brightest
features ($T_{mb} > 1$\,K).  In total, we tabulate measurements for
140 lines brighter than 1\,K.  These measurements represent only a
small fraction of 727 spectral features detected in the data.  The
species associate with each of the lines in Table~4 were identified
using the spectral line catalogs given within
Splatalogue\footnote{http::/www.splatalogue.net/} which is based on
several spectroscoptic databases (M\"{u}ller et al. 2005; Prickett et
al. 1998; Lovas \& Dragoset 2004).  For each line, we list the most
likely species.  For the bright lines listed in Table~4, only one
feature was not identified.  This transition was measured at $T_{mb} =
3.9$\,K at an observed frequency of 81.756\,GHz.  Species with
cataloged transitions closest in the frequency to this unidentified
line include c-HCOOH (formic acid), H$_2$NCH$_2$CN
(aminoacetonitrile), CH$_3$COCH$_3$ (acetone), and CH$_2$ND, but none
of these possible identifications seem particularly compelling due to
significant velocity offsets or the lack of other detected transitions
associated with the species.

The vast majority of the spectral lines detected are associated with
the cooler molecular gas regions consistent with velocities of $\vlsr
=8.8 \kps$ and widths of about $5 \kps$, or with the ``hot core''
region with $\vlsr \approx 5 \kps$ and widths of about $10 \kps$
(Sutton et al. 1985).  The SO$_2$ features show an additional broad
component that may be associated with a strong outflow in the system
(e.g., Wright et al. 1996).

The spectral lines observed with the 100-meter GBT are typically a few
to more than 10 times brighter than those previously measured by the
much smaller NRAO 36 Foot telescope (Turner 1989), and the brightest
lines (e.g., SiO and HCN) observed with the GBT also tend to be
brighter than those observed with the IRAM 30m (Carter et al. 2012).
The variations in line strengths from different telescopes highlight
the effects of beam dilution and filling factor effects where the
observed intensity from small emission regions decreases with an
increasing telescope beam size.  However, many species also show good
agreement in their observed line strengths.  We find similar line
strengths within measurement errors between the tabulated 30m
measurements (Esplugues et al. 2013; Tercero et al. 2010) and those
presented in Table~4 from the GBT for the SO, $^{34}$SO, SO$_2$, and
OCS transitions.  These results suggest that the emission from these
species are extended over both the 30m and GBT beams.  The
interpretation of line strengths is also complicated due to the
multiple overlapping components of the Orion complex (extended ridge,
compact ridge, plateau, and hot core; e.g., Tercero et al. 2010).
Telescopes with different beam sizes will not observe the exact same
portion of the Orion complex in a single pointing.  High-spatial
resolution mapping observations would be useful in disentangling the
relative line strengths from the different components in Orion given
that the components also have similar velocities (e.g., Widicus
Weaver \& Friedel 2012; Brouillet et al. 2013).

\section{Concluding Remarks}

In this paper we present the 67--93.6 GHz spectral-line survey of
Orion-KL taken with the GBT.  The full spectrum is available
electronically.  We tabulate spectral-line measurements for features
brighter than $T_{mb} > 1$\,K, but features as weak as 10\,mK are
identifiable in the data.  The survey targeted the bright continuum
peak of Orion-KL.   Many molecules are known to have strong emission
elsewhere within Orion (e.g., Peng et al. 2013), and the GBT beam of
$10\arcsec$ would be well-suited for mapping the distribution of
molecular emission throughout the Orion complex.

Given the large collecting area of the GBT, the improved accuracy of
the surface, and the excellent performance of the optimized LNAs, the
GBT 4\,mm system is the most sensitive facility in the world for this
band.  Within the over-lapping frequency range of ALMA band-3, the GBT
system can achieve better system temperatures in spite of the higher
opacity associated with the lower elevation of the Green Bank site.
With the multi-pixel instruments currently under development (e.g.,
Argus, Sieth et al. 2014; Mustang-2, Dicker et al. 2014), the mapping
speed within this band using the GBT would be unparalleled.

\acknowledgments

We thank our colleagues at Green Bank and at the NRAO Central
Development Lab who have made these observations possible.  We thank
Simon Dicker and Brian Mason for sharing the GBT Mustang data shown in
Figure~1.  The National Radio Astronomy Observatory is a facility of
the National Science Foundation operated under cooperative agreement
by Associated Universities, Inc.  {\it Facility:} \facility{GBT}

\begin{figure}[tbh]
\plotone{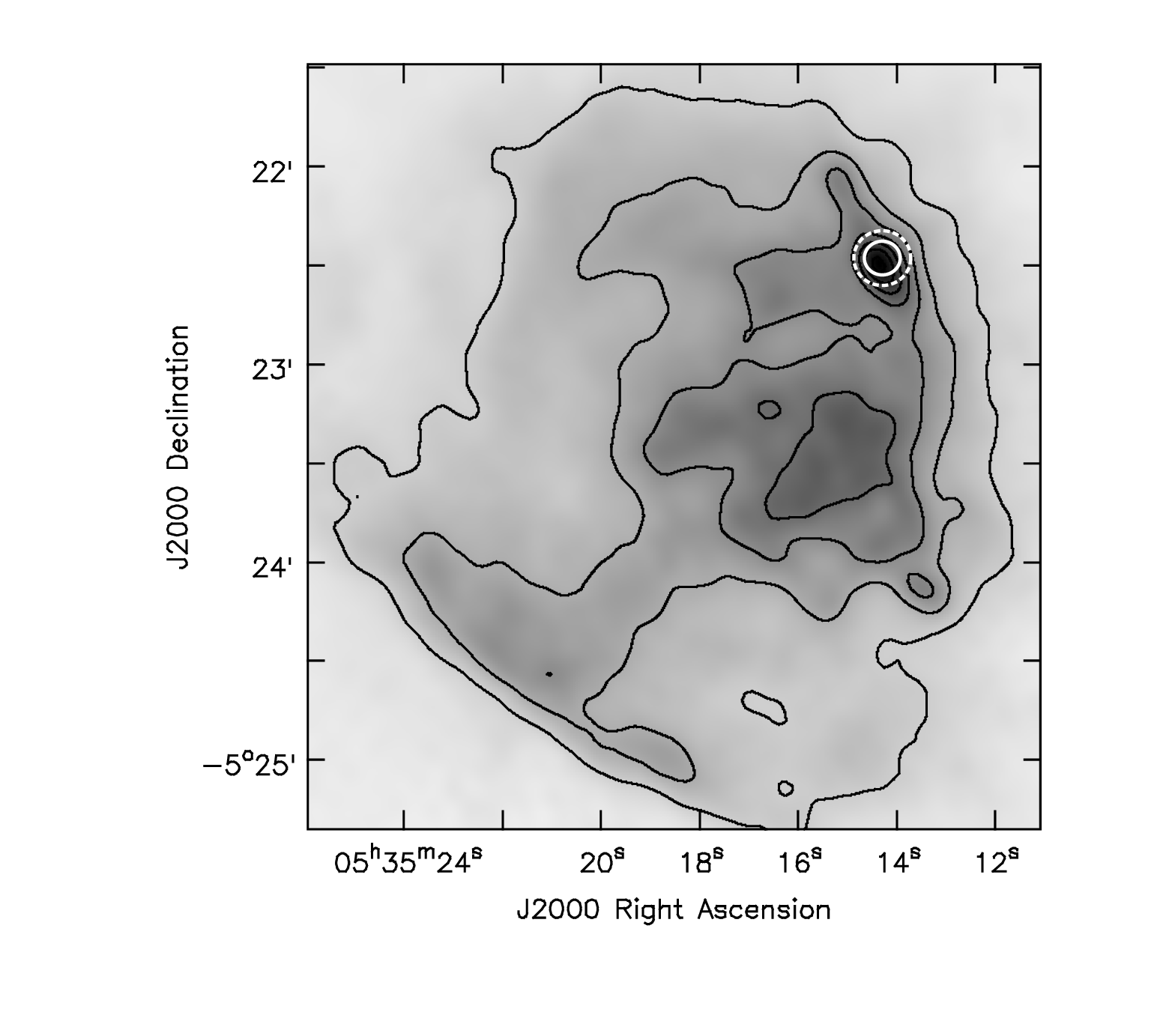}
\vspace*{-1mm}
\caption{The GBT Mustang 90\,GHz continuum image of Orion shown as
  grey-scale and contours (Dicker et al. 2009).  The contours start at
  0.2\,Jy and are incremented by 0.2\,Jy.  The location of the nominal
  GBT beam at 77\,GHz of the spectral-line observations is shown by
  the solid white circle, which corresponds to the bright continuum
  position of Orion-KL.  The dotted white circle shows the possible
  area observed including the pointing errors and the variation of the
  beam size during the daytime observations.}

\end{figure}

\begin{figure}[tbh]
\plotone{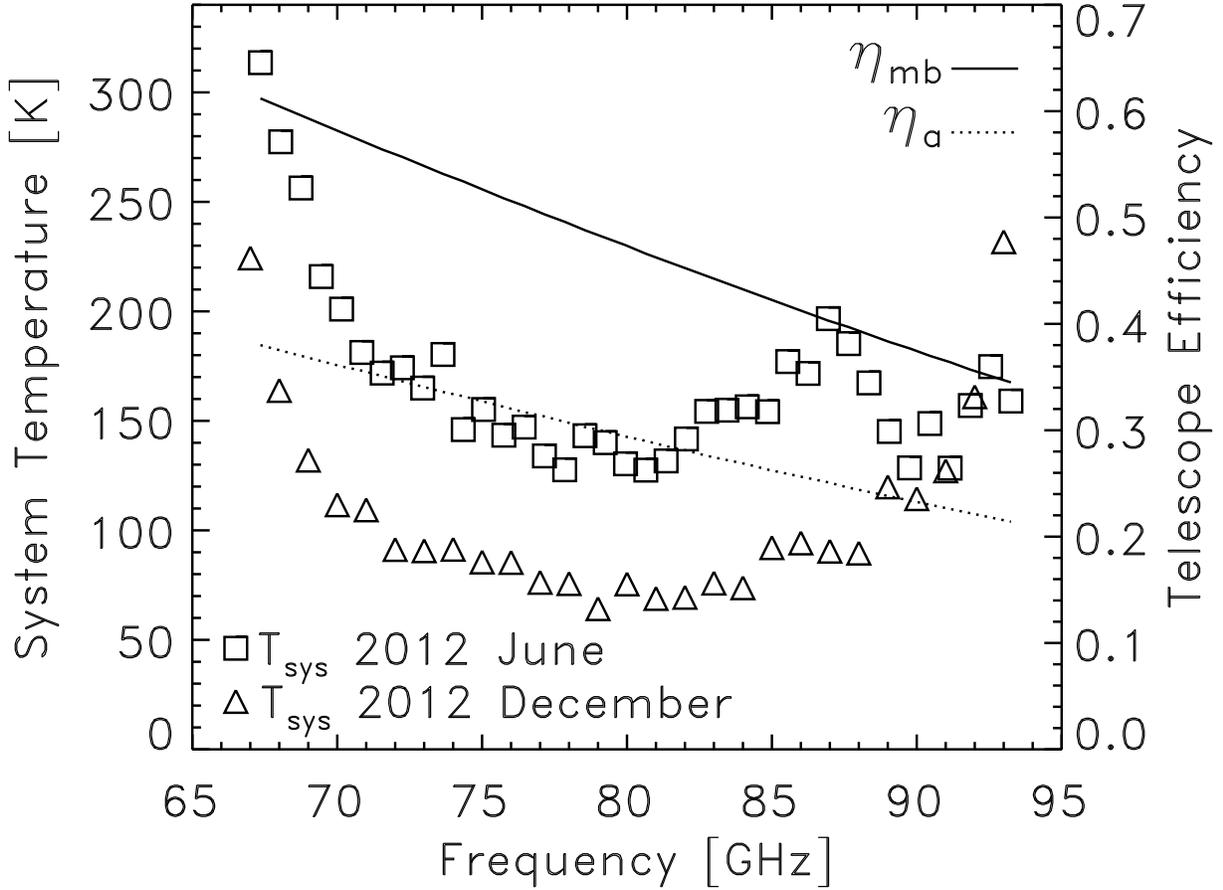}
\vspace*{-1mm}
\caption{The measured system temperatures for the Orion-KL data are
  shown by squares (2012 June), and the improved system temperatures
  with the upgraded cold amplifies are shown by triangles (2012
  December).  Both data sets were taken in similar atmospheric
  opacity.  The telescope efficiency scale is shown to the right.  The
  measured main-beam efficiency ($\eta_{mb}$) and aperture efficiency
  ($\eta_{a}$) are shown as a function of frequency by the solid and
  dotted lines, respectively.}

\end{figure}

\begin{figure}[tbh]
\includegraphics[width=0.75\textwidth,angle=90]{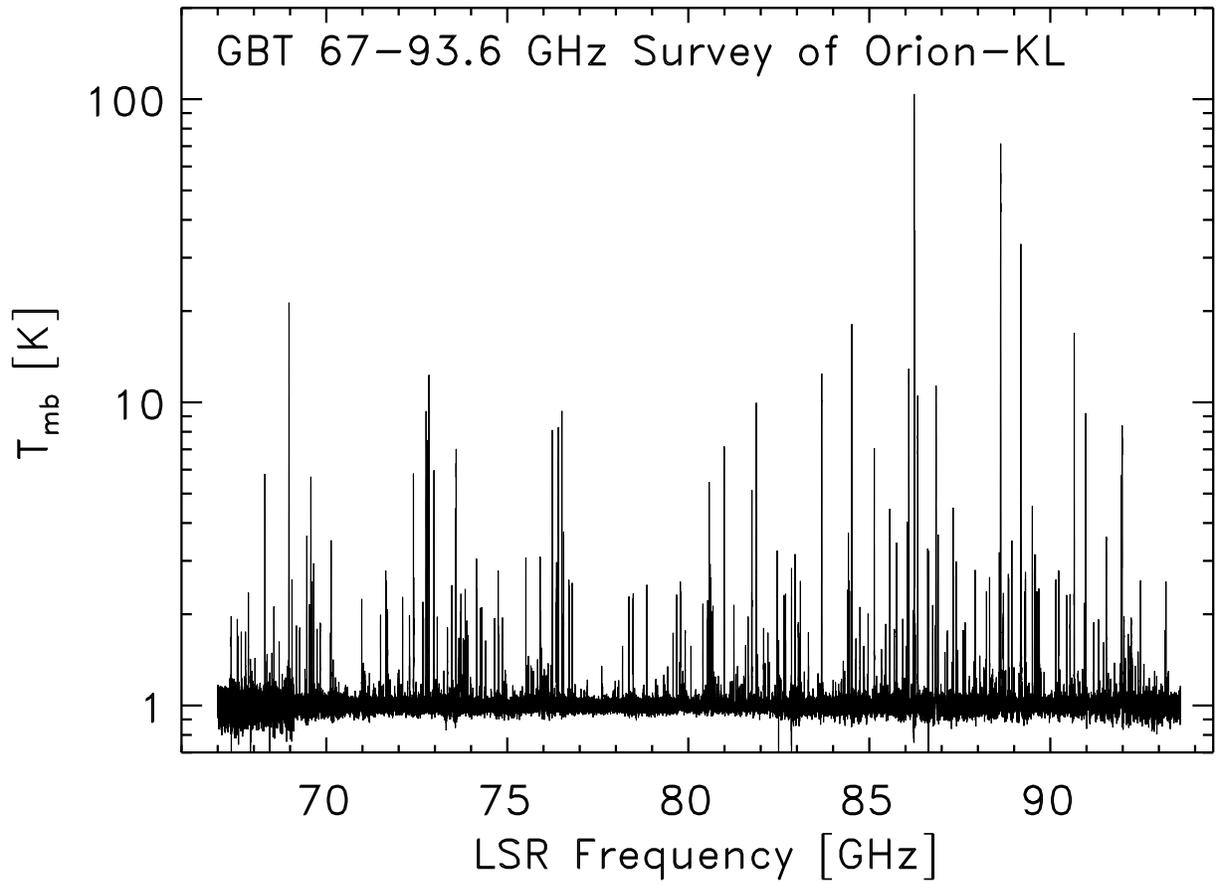}
\vspace*{-1mm}
\caption{The 67--93.6 GHz spectrum of Orion-KL taken with the GBT.
  The data have been baseline subtracted, but the spectrum has been
  shifted by $+1$\,K for plotting on a logarithm scale.}

\end{figure}

\begin{figure}[tbh]
\plotone{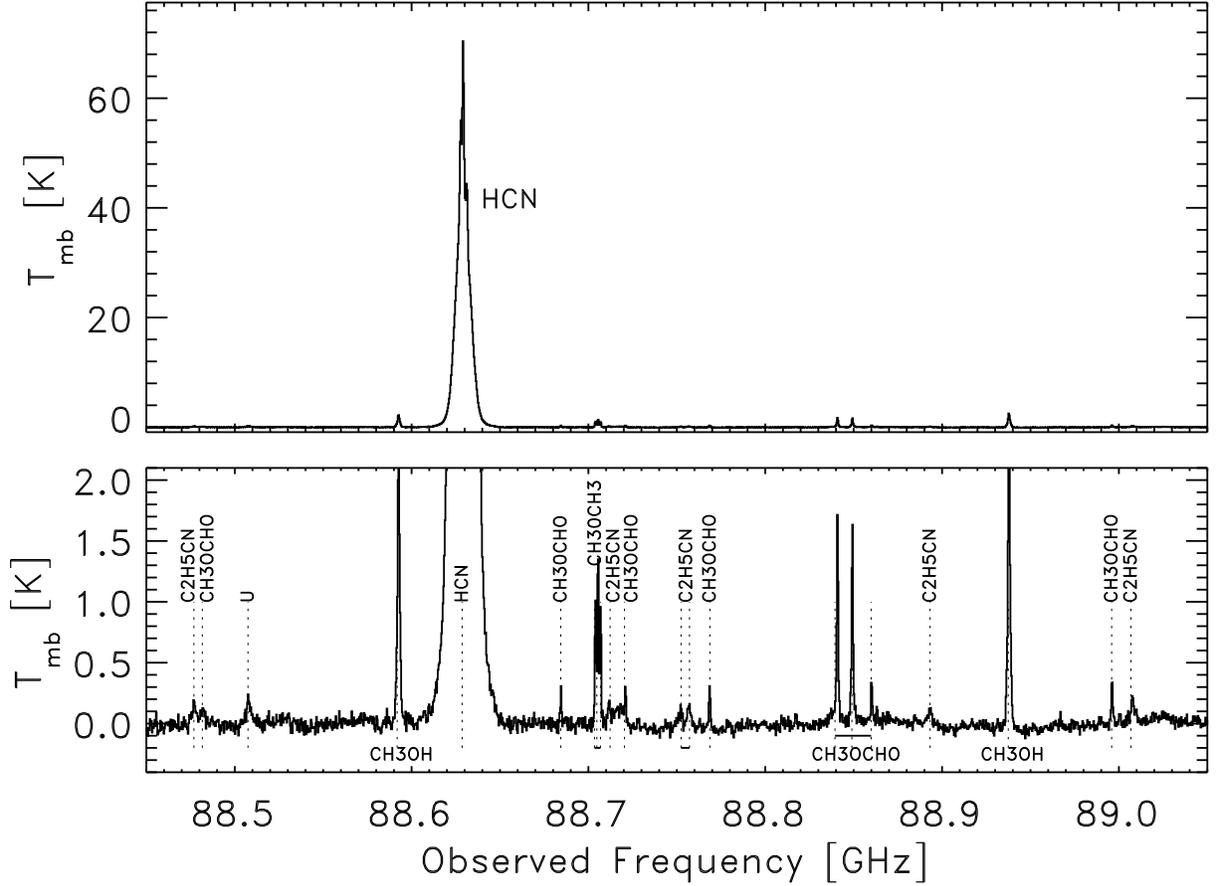}
\vspace*{-1mm}
\caption{An example portion of the survey that includes the strong
  HCN(1-0) emission line.  The top panel shows the full data range,
  while the bottom panel shows the same data zoomed-in along the
  y-axis to highlight the visibility of weak features.  Spectral
  features are marked by the dotted-lines and labeled by species.  The
  dotted-lines of neighboring transitions from the same species are
  connected by a solid line below the spectrum for clarity.  The
  feature labeled with a ``U'' is unidentified.  The full survey
  plotted similarly with overlapping spectra separated by 0.5 GHz is
  provided as supplemental online material (Fig. 5).}

\end{figure}

\begin{table}
\caption{Observational Parameters}
{\small
\begin{tabular}{ccccc}
\tableline\tableline
Frequency  & T(on) & $T({\rm sys})$ & $\eta_{mb}$ & $\sigma$ \\
 GHz      &  min & K &            & mK \\
(1)        & (2)   & (3)            &  (4)       & (5)\\
\tableline
67.35 &  24.55 &   314 & 0.61 &  77.0 \\
68.05 &  17.85 &   277 & 0.60 &  61.7 \\
68.75 &  19.80 &   256 & 0.60 &  71.2 \\
69.45 &  19.80 &   216 & 0.59 &  43.9 \\
70.15 &  19.80 &   201 & 0.58 &  36.7 \\
70.85 &  20.75 &   181 & 0.57 &  33.0 \\
71.55 &  20.80 &   172 & 0.56 &  32.0 \\
72.25 &  20.80 &   174 & 0.56 &  28.4 \\
72.95 &  20.80 &   165 & 0.55 &  33.5 \\
73.65 &  19.60 &   180 & 0.54 &  33.9 \\
74.35 &  19.80 &   146 & 0.53 &  26.8 \\
75.05 &  19.80 &   155 & 0.53 &  29.5 \\
75.75 &  19.80 &   143 & 0.52 &  26.9 \\
76.45 &  19.80 &   147 & 0.51 &  32.5 \\
77.15 &  18.80 &   134 & 0.50 &  27.6 \\
77.85 &  18.80 &   128 & 0.50 &  25.0 \\
78.55 &  18.80 &   143 & 0.49 &  30.9 \\
79.25 &  18.80 &   140 & 0.48 &  31.3 \\
79.95 &  19.80 &   129 & 0.47 &  24.6 \\
80.65 &  19.80 &   126 & 0.47 &  22.6 \\
81.35 &  19.80 &   130 & 0.46 &  26.7 \\
82.05 &  19.90 &   142 & 0.45 &  33.0 \\
82.75 &  18.80 &   154 & 0.45 &  40.8 \\
83.45 &  18.80 &   155 & 0.44 &  31.9 \\
84.15 &  18.80 &   156 & 0.43 &  33.0 \\
84.85 &  18.80 &   154 & 0.42 &  39.4 \\
85.55 &  19.80 &   177 & 0.42 &  36.9 \\
86.25 &  19.80 &   172 & 0.41 &  45.9 \\
86.95 &  19.80 &   197 & 0.40 &  43.9 \\
87.65 &  19.80 &   185 & 0.40 &  38.5 \\
88.35 &  19.80 &   167 & 0.39 &  37.9 \\
89.05 &  19.80 &   145 & 0.38 &  38.2 \\
89.75 &  19.80 &   129 & 0.38 &  31.9 \\
90.45 &  19.80 &   149 & 0.37 &  36.1 \\
91.15 &  19.80 &   128 & 0.36 &  45.3 \\
91.85 &  19.80 &   157 & 0.36 &  50.2 \\
92.55 &  19.80 &   175 & 0.35 &  59.2 \\
93.25 &  19.65 &   159 & 0.34 &  52.3 \\
  
\tableline
\end{tabular}

\tablenotetext{}{(1) The center frequency of each spectral window.  (2)
  The total on-source integration time after data editting. (3) The
  effective system temperature.
  (4) The main-beam efficiency (Eqn. 4).  (5) The rms
  noise level in units of $T_{mb}$ per 0.39\,MHz channel.}  }
\end{table}

\begin{table}
\caption{The GBT 67 -- 93.6 GHz Spectrum of Orion-KL}
\begin{tabular}{rr}
\tableline\tableline
LSR Frequency  & $T_{mb}$ \\
 GHz      & K  \\
(1)       & (2) \\
\tableline
   66.998244 &  -0.008832\\
   66.998634 &   0.123508\\
   66.999025 &   0.079210\\
   66.999416 &   0.023435\\
   66.999806 &  -0.065857\\
   67.000197 &   0.092554 \\
   \dotfill  & \dotfill \\
\tableline
\end{tabular}

\tablenotetext{}{(1) The observed LSR frequency in GHz.  (2)
  The main-beam telescope temperature in Kelvin corrected for
  atmospheric and telescope losses.  The table only shows the first 6
  rows to provide an example of content.  The complete data set comprising
  68,096 rows is published electronically.}
\end{table}

\begin{table}
\caption{Data Artifacts}
\begin{tabular}{ll}
\tableline\tableline
Frequency  & Feature \\
 GHz      &    \\
\tableline
67.370   & Atmospheric O$_2$\\
67.901   & Atmospheric O$_2$\\
68.431   & Atmospheric O$_2$\\
82.495  & RFI\\
82.851   & RFI\\
86.629   & RFI\\
\tableline
\end{tabular}
\end{table}

\begin{table}
\caption{Bright Line Measurements and Identification}
\begin{tabular}{cccc}
\tableline\tableline
Frequency &  Line Peak & FWHM & Identified Species \\
 MHz      &   K        & MHz  &                     \\
 (1)      &   (2)      & (3)  &  (4) \\
\tableline
  67847.503$\pm$0.074&   1.30$\pm$0.09&  2.11$\pm$0.23&  SO2\\
  68303.919$\pm$ 0.035&   4.68$\pm$0.41&  0.78$\pm$0.08&  CH3OH\\
  68553.163$\pm$0.033&   1.10$\pm$0.09&  0.83$\pm$0.08&  CH3OH\\  
68970.754$\pm$0.075&  19.20$\pm$0.71&  3.65$\pm$0.24&  SO2\\
\dotfill & \dotfill & \dotfill & \dotfill \\
\tableline
\end{tabular}
\tablenotetext{}{Results of fitting the brightest lines
  ($T_{mb} >1$\,K) with a Gaussian.  (1) The center line frequency in
  MHz.  (2) The fitted line peak in $T_{mb}$ [K].  (3) The measured FWHM of the
  line in MHz. (4) The most likely species associated with the
  line. The table only shows the first 4 rows to provide an example of
  content.  The complete table comprising 140 rows is published electronically.}
\end{table}

\end{document}